\input{aipcheck.tex}

\documentclass[aipproc]{aipproc}

\layoutstyle{8x11double}

\SetInternalRegister\hbadness{8000} 

\newcommand\doingARLO[2][]{%
  \ifx\mmref\undefined #1\else #2\fi
}

\begin{document}

\title 
      [GRB Prompt Emission.]
      {Towards an Understanding of Prompt GRB Emission.}

\classification{43.35.Ei, 78.60.Mq}
\keywords{Document processing, Class file writing, \LaTeXe{}}

\author{Nicole M. Lloyd-Ronning}{
  address={Canadian Institute for Theoretical Astrophysics},
  email={lloyd@cita.utoronto.ca},
}

\copyrightyear  {2001}

\begin{abstract}
  We discuss the prompt emission of Gamma-Ray Bursts in different
  spectral energy bands.  First,
  we suggest that a three-part synchrotron emission model \cite{lp00, lp01} is a good
  description of the $\sim 20$ keV - $1$ MeV gamma-ray emission of GRBs.  We show that this
  model provides excellent fits to the data and naturally explains
  the observed global correlations between spectral parameters.
  In particular, we show there exists a negative correlation between
   between the peak
   of the $\nu F_{\nu}$ spectrum, $E_{p}$, and the low
   energy photon index $\alpha$ for bursts with
   $-2/3 < \alpha < 0$, and suggest that this correlation is due
   to the mechanism
   responsible for producing $\alpha$'s above the 
   value of $-2/3$ - namely, a decreasing mean pitch angle
   of the electrons. 
   We then discuss the physical origin of the increasing number of GRBs
   that are observed to peak in the X-ray energy band ($\sim 5-40$ keV).
    Although either a cosmological (i.e.
   high redshift) or intrinsic interpretation for the low values of
   $E_{p}$ is viable at this point,  the data appear to suggest that intrinsic
   effects are playing the dominant role.  
   Finally, we briefly comment on the prompt GRB optical emission ($\sim$ eV) and very high
   energy emission ($>10$ MeV), and how
   these spectral bands may be used to place additional constraints on the physics of 
   gamma-ray bursts.
\end{abstract}

\date{\today}

\maketitle

\section{Introduction}
   
  Just as broadband observations  were (and are) necessary
  to establish
  the general external shock paradigm of the GRB afterglow,
  so are broadband data necessary to fully understand the physics
  of the prompt GRB emission.  The prompt ($\sim$ first one hundred
  seconds or so) emission  gives us a unique
  opportunity to explore the physics of internal - and possibly external - 
  shocks, address fundamental issues associated with the
  behavior of relativistic plasmas, particle acceleration
  and turbulence, and possibly even gain insight into the GRB progenitor itself.
  In this paper, I will present some recent advancements in our understanding
  of GRB prompt emission,
  and discuss important remaining questions associated with the initial
  phases of a burst.

\section{Prompt Gamma-Ray Emission}

Most GRB spectra in the energy range ($\sim 20$ keV - $\sim$ MeV) are well
described by a so-called
   Band spectrum \cite{ba93}.  This model is
   essentially a smoothly broken power with a low energy photon spectral
   index $\alpha$, a high energy photon index $\beta$, and a break
   energy $E_{p}$.  There have been
   some attempts to explain or interpret  the global properties
   of these spectral parameters in
   terms of a physical model \cite{tav96, pre96, ghi99}, with
   inconclusive results.  In particular, it was thought from very early on
   that the radiation mechanism responsible for GRBs (in this energy range) is
   synchrotron emission \cite{kat94}.  However, problems with this
   model - particularly with the behavior of the low energy $\gamma$-ray
   spectrum - were quickly brought to light. First, it was noted \cite{pre96}
   that a number of
   bursts have a low energy spectral index $\alpha$ that falls above the
   so-called ``line-of-death'' value of $-2/3$, the asymptotic limit
   of an instantaneous (i.e. non-cooling)
   optically thin synchrotron spectrum from an isotropic, power-law electron
   distribution with some minimum cutoff energy.
     Even more fatal for the standard synchrotron picture, it was pointed out
   \cite{ghi99} that if particles
   are injected at once and then left to radiate, the synchrotron cooling times
   are very short (much less than the detector integration time).  In this
   case, all of the particles cool and the
   electron spectrum becomes a soft power law (of
   index $-2$) at low energies. As a result,  the asymptotic (upper) 
   limit of the low energy photon spectrum is $\alpha = -3/2$. Nearly {\em all} GRBs
   have observed values of $\alpha$ that fall above this limit!  
   However, as we will show below, when some of the simplifying assumptions
   in these models are modified, synchrotron emission in fact does a
   very good job of explaining the observed data.

   \subsection{A Three-part Synchrotron Model}
Our model (described in more detail in \cite{lp00} and \cite{lp01}) 
modifies the usual simple
    picture of optically thin synchrotron emission from a power
    law distribution of electrons with a sharp low energy cutoff,
    by accounting for: 1) the possibility of a smooth cutoff to the low 
    energy electron
    distribution, 2) radiation from an anisotropic
    electron distribution with
    a small mean pitch angle, 3) synchrotron self-absorption, and
    4) the important instrumental effect in which the value
    of the fitted
    parameter $\alpha$ decreases as $E_{p}$ approaches 
    the lower edge of the BATSE
    window.  We have envisioned a realistic scenario in which particle
    acceleration and synchrotron losses occur continually and simultaneously
    behind each internal shock, with the characteristic acceleration
    time shorter than the loss time;
    this means that synchrotron loss effects are only
    evident in the particle distribution spectrum at energies much
    larger than what is relevant for our discussion here (at energies where the inequality
    is reversed and the loss time becomes shorter than the acceleration
    time).  The emission in this model can be characterized by three distinct
    regimes. 
  \begin{figure}
  \includegraphics[height=.35\textheight]{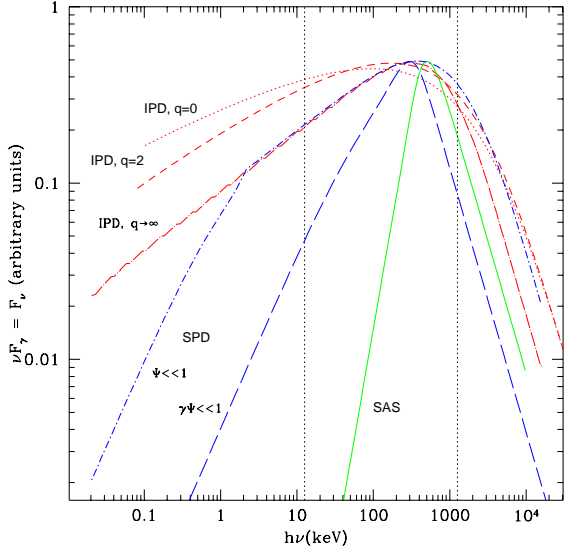} 
  \caption{Synchrotron spectra in the various emission regimes
  (see text for explanation).  The dotted vertical lines represent the approximate width
  of the BATSE spectral window.}
\end{figure}

1) \underline{\bf IPD.} This is the familiar optically thin synchrotron emission 
from a 
power law  electron energy spectrum, with an isotropic pitch
angle distribution. In contrast to 
most analyses, however, here we consider an electron distribution 
 with a smooth low energy cutoff: $N(\gamma) \propto
\frac{(\gamma/\gamma_{m})^{q}}{1+(\gamma/\gamma_{m})^{p+q}}$. 
Note that for high energies ($\gamma > \gamma_{m}$),
 the spectrum goes as $\gamma^{-p}$, while
for low energies ($\gamma < \gamma_{m}$), the spectrum goes
as $\gamma^{q}$. Hence, $q$ denotes the 
steepness of the
electron low energy cutoff (note that an actual ``cutoff'', in the sense
that $N(\gamma) \rightarrow 0$ as
$\gamma \rightarrow 0$, requires $q > 0$).  The
asymptotic behavior of the synchrotron (photon number)
spectrum for $q > -1/3$ is:
\begin{equation}\label{equation} F_{\gamma} = \cases{\nu^{-2/3}&
$\nu \ll \nu_{m} = \frac{2}{3} \nu_{B} {\rm sin}\Psi \gamma_{m}^{2}$\cr
\nu^{-(p+1)/2}&  $\nu \gg \nu_{m}$}
\end{equation} 
where $F_{\gamma}$ is the photon flux, $\Psi$ is the electron pitch angle,
and $\nu_{B} = \frac{eB}{m_{e}c}$
where $B$ is the magnetic field.  Note that  the
peak of the $\nu F_{\nu}$ spectrum will occur at
$E_{p} \propto \nu_{m} \propto B{\rm sin}\Psi 
\gamma_{m}^{2}$, and that the aymptotic low energy index below this break 
is $\alpha = -2/3$.  

In this regime, \underline{we expect a positive 
 correlation} between $E_{p}$ and $\alpha$ due to instrumental
 effects alone.  If $E_{p}$ is close to
 the edge of the BATSE window, the low energy photon index may not 
 yet have reached
 its asymptotic value and a smaller (or softer) value of $\alpha$
 (relative to the asymptotic value)
 will be determined.  A smooth cutoff to the electron energy
 distribution will exacerbate this effect because for a smoother
 cutoff (or a smaller value of $q$), the asymptote is reached
 at lower energies relative to $E_{p}$. 
  Note that a dispersion in the
 smoothness of the low energy cutoff will tend to wash this correlation
 out to some degree, as seen in Figure 4 of \cite{lp00}. 
 For the cases of small pitch angle radiation and the self-absorbed
 spectrum (see below), this instrumental effect will be weaker because the low
 energy asymptotes are reached more quickly (i.e. at energies closer
 to $E_{p}$) than
 for the isotropic optically thin case (see Figure 1).
 
2) \underline{\bf SPD.} 
 This type of synchrotron spectrum results from optically thin 
synchrotron emission by electrons with a mean pitch angle $\Psi \ll 1$;
the analysis of synchrotron radiation in this regime was first
done by \cite{ep73}.
For low density, high magnetic field plasmas expected in GRBs, the Alfv\'en phase 
velocity is greater than the speed of light and (therefore) the speed of the 
relativistic particles under consideration here. In this case, the pitch angle diffusion 
rate of the electrons interacting with plasma turbulence is 
smaller than the acceleration rate; consequently, the accelerated electrons
could maintain a  highly 
anisotropic distribution as required in the small pitch angle model. The shape of
this spectrum depends on just how small the pitch angle is. For
$\Psi \ll 1$, but $\Psi\gamma_{m} \sim 1$, we have:
\begin{equation}\label{equation} F_{\gamma} = \cases{\nu^{0}&
$\nu \ll \nu_{s} = \frac{2}{3}\nu_{B}/(\gamma_{m}\Psi^{2})$\cr
\nu^{-2/3}&  $\nu_{s} \ll \nu \ll \nu_{m}$\cr
\nu^{-(p+1)/2}&  $\nu_{m} \gg \nu$}
\end{equation}
There are two breaks in this spectrum - one
at $\nu_{m}$ and one at $\nu_{s}$.  Because the Band spectrum can
only accommodate one break, spectral fits to
this model will put the parameter $E_{p}$ at one
or the other of these two breaks, but most likely at $\nu_{m}$ 
 because
for $p > 5/3$ (or for  high energy photon
index $\beta < -4/3$ which is the case for most bursts), the break across $\nu_{m}$ is more
pronounced than across $\nu_{s}$.  In this case, the low energy photon
index $\alpha$ will fall somewhere between $-2/3$ and $0$. 

 However, as
the pitch angle $\Psi$ decreases such that $\Psi \ll 1/\gamma_{m}$, then
the $\nu^{-2/3}$ portion of the spectrum disappears, and only the
$\nu^{0}$ portion is left.  In this case we have:
\begin{equation}\label{equation} F_{\gamma} = \cases{\nu^{0}&
$\nu \ll \nu_{s} =\frac{4}{3}\nu_{B}\gamma_{m}$\cr
\nu^{-(p+1)/2}&  $\nu_{s} \gg \nu$,}
\end{equation}
where $E_{p} \propto B\gamma_{m}$ (see \cite{ep73} for a more
detailed description of the behavior of the spectrum in this regime).

Here, we expect evidence of 
 a \underline{negative correlation} between $E_{p}$ and $\alpha$ as we transition
 from the IPD to the SPD regime, i.e.  for $-2/3<\alpha<0$.  In this case, the pitch angle  
 decreases so that $E_{p} \propto {\rm sin}\Psi$ decreases, if all other
 physical parameters ($B$ and $\gamma_{m}$) remain constant. In
 addition, as we
go from the small pitch angle regime, $\Psi \gamma_{m} \sim 1$ ($\Psi \ll 1$), to
the very small pitch angle regime,  
to $\Psi \gamma_{m}
\ll 1$, the $\nu^{-2/3}$ portion of the
spectrum disappears, and we are left with only the $\nu^{0}$ portion.
In other words, as the mean of 
the pitch angle distribution decreases to very small values,  $E_{p}$ decreases
and the value of $\alpha$ decreases from $-2/3$ to $0$.
   This negative correlation will compete with the
 positive instrumental correlation mentioned above.
 
3) \underline{\bf SAS.} If the magnetic field and density are such that the medium becomes
optically thick to the synchrotron photons with frequency $\nu < \nu_{a}$,
then, for $\nu_{a} < \nu_{m}$,
 we have the following spectrum:
\begin{equation}\label{equation} F_{\gamma} = \cases{\nu^{1}&
$\nu \ll \nu_{a}, $\cr
\nu^{-2/3}&  $\nu_{a} \ll \nu \ll \nu_{m}$,\cr
\nu^{-(p+1)/2}&  $\nu_{m} \gg \nu$}.
\end{equation}
In that case,  $E_{p} \propto \nu_{a} \sim 
10 (nl)^{3/5}B^{2/5}\gamma_{m}^{-8/5}\Gamma^{9/5}$ Hz,
 where $l$ and $n$ are the path length and particle density in the co-moving
 frame, and
 we have assumed an electron energy distribution index $p=2$.  
  For $\nu_{a} >
 \nu_{m}$ we
just have one break at $\nu_{a}$ with a low energy photon index
of $\alpha = 3/2$ (in both the isotropic and small pitch angle cases).  The possibility of
self-absorption in GRBs is a controversial issue.
We have shown \cite{lp00} that there 
are  bursts for which a self-absorbed spectrum is a better fit than an 
optically thin one.  We also found that in these cases, the absorption
frequency tends to be near the lower edge of the BATSE
window.  In addition to this,
Strohmeyer et al. \cite{str98} found that a number
of bursts observed by GINGA with $E_{p}$'s in the
range 2 to 100 keV have steep ($\alpha \sim  1$) low energy spectral indices
consistent with a self-absorbed spectrum.
This raises interesting questions about the 
 physics of the ambient plasma, because self-absorption in a GRB
requires fairly large ($\sim 10^{8} G$) magnetic fields 
and particle densities ($\sim 10^{8} cm^{-3}$).  
  The physical processes  required to achieve these conditions
 will need to be theoretically
 established if the data prove self-absorption 
 to be a viable model.

\subsection{How the Data Stand Up}
\underline{\bf Spectral Fits:}
   We fit each spectrum to all 3 emission
 scenarios and then evaluate the fits based on their values of
 a reduced $\chi^{2}$. 
  In Figure 2, we show examples of spectral fits in each emission regime.
 Each fit is taken at a time during the burst spectral evolution when
 the $\alpha$ parameter corresponded to the respective model.  For example,
 in the top panel - burst 1663 - the spectrum is from a time when $\alpha \sim
 -2/3$,
 while in the middle panel - burst 105 - the spectrum is from a time in
 the profile when $\alpha = 0$.  Similarly, for the bottom panel, this spectrum
 corresponds to a time when $\alpha=1$.  The reduced $\chi^{2}$ are $0.34$,
 $0.33$, and $0.50$ for the top, middle and bottom panels respectively.
  In general the best model turns
 out to correspond to
 the emission regime suggested by Band's $\alpha$ values, which confirms
 our proposed method of  physically interpretating  Band fits based on
 the bursts' low energy photon index  
 (for example, an IPD fit to the spectrum of burst 105
 gave a $\chi^{2} > 1$ compared to the $\chi^{2} = 0.33$ for an SPD fit).
\begin{figure}[t]
  \includegraphics[height=.35\textheight]{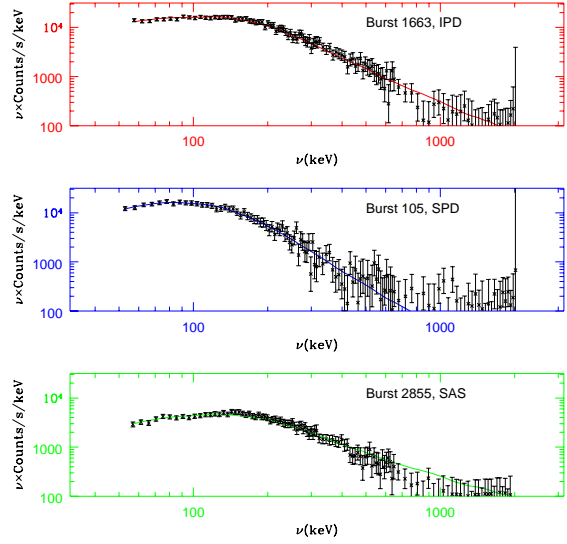}
  \caption{Three time resolved spectra fit to our generalized
  synchrotron model.  The best fit emission regime corresponds to
  what is predicted by the low energy Band parameter $\alpha$.}
  \end{figure}

\underline{\bf Global $\alpha$ Distribution:} 
  As discussed above, the low energy photon index
   $\alpha$ is the best parameter for distinguishing between
the various synchrotron regimes.
 Figure 3 shows a histogram of $\alpha$ (taken from 2,026
  time resolved spectra with Band spectral fits, from \cite{pre99}) with each regime 
  clearly marked. 
Note that there are a significant number of spectra in the SPD regime.

 \begin{figure}[t]
  \includegraphics[height=.35\textheight]{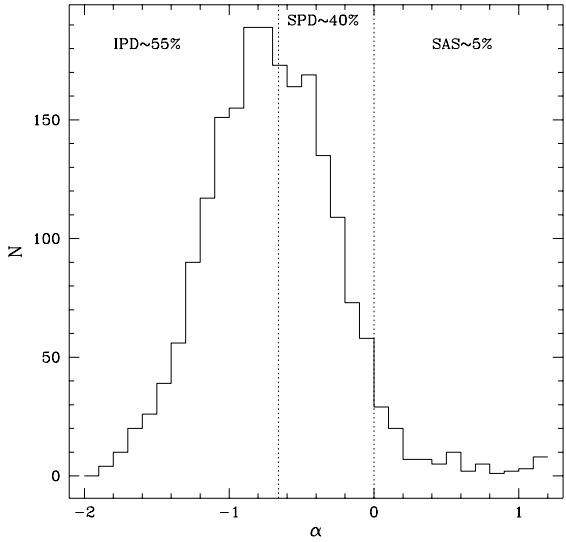}
  \caption{Histogram of the low energy spectral index $\alpha$,
  with each emission regime delineated.  The data are from
  Preece et al., 1999.}
  \end{figure}
  
 Although the error bars on $\alpha$ can make some
difference in the numbers of spectra in each regime, 
this does not affect  the
qualitative nature of our conclusions below (see \cite{lp01} for a more
detailed discussion on how the error on $\alpha$ affects this distribution).
 We now discuss the correlations present in the data and their consistency with
 what we expect in
 the context of the three synchrotron emission scenarios.
 
 \underline{\bf Observed Correlations} 
 Figure 4 shows the binned average correlation between $E_{p}$ and $\alpha$
present in $\sim 2000$ time resolved spectra from \cite{pre99}. We have sorted $\alpha$
in ascending order and binned the data every 100 points (the horizontal
error bars indicate the size of the bins).  We then computed the
average  $E_{p}$ for these 100 points.  The most intriguing result
is that the correlation appears to be positive for $\alpha < -0.7$
and negative for $\alpha > -0.7$.
Performing a Kendell's $\tau$ test
on all of the (unbinned) data, we find a $9 \sigma$ {\em positive correlation
between $\alpha$ and $E_{p}$ in the IPD regime}.  To account for both
the error in $\alpha$ and $E_{p}$, we have performed this test on
all permutations of correlations between the lower and upper values of $\alpha$ (from
the $1\sigma$ error bars) with the lower and upper values of $E_{p}$. In
addition we have averaged the value of the correlation
statistic $\tau$ from 
 100 sets of data, in which - for each data point - 
$\alpha$ and $E_{p}$ are drawn from Gaussian distributions
with means equal to the parameter
 values given in the catalog and standard deviations corresponding to
the error bars. {In all cases, we find a highly
significant ($>6\sigma$) correlation}.  
The positive correlation between $\alpha$ and
$E_{p}$ in the IPD regime
can be simply understood by the instrumental effect
discussed in the previous section.

On the other hand, we find a $4 \sigma$ {\em negative correlation
between $\alpha$ and $E_{p}$ in the SPD regime}.  Again, to account for  
the error in both $\alpha$ and $E_{p}$, we have performed this test on
all permutations of correlations between the lower and upper values of $\alpha$ (from
the $1\sigma$ error bars) with the lower and upper values of $E_{p}$, and have also
averaged the $\tau$ value from 100 sets of data drawn from distributions
based on the existing data, according to the prescription described in the above paragraph. 
 {In all of these cases, we find a  
significant ($>3\sigma$) negative correlation}.
As mentioned above, this type of correlation is natural in the
small pitch angle regime, as a result of a decreasing average pitch
angle in the electron distribution.

The observed trends are 
 consistent with what is expected from our model in each
 emission scenario, and tell us something important about the role various effects
 play in the correlations.
For example,
 the dashed line in Figure 4 shows how $E_{p}$ should change as a function of $\alpha$ in the
BATSE spectral window,
if only the mean of the electron pitch angle $\Psi$ changes (all
other parameters such as $B$ and $\gamma_{m}$ remaining constant).
The fact that the observed correlation is weaker could be due to a number of
different
physical effects.\footnote{The positive instrumental correlation discussed in the
previous section will also play
a small role in reducing the strength of the negative correlation.} 
Of course, we expect that the correlation will be 
washed out to some degree by dispersion in the 
 intrinsic values of
$\Psi$ and $\gamma_{m}$, as well as
variation in the magnetic field from burst to burst.  It is also possible that
the minimum electron Lorentz factor or the magnetic
field of the electrons
{\em increases} as we transition to a physical regime in which electrons
are accelerated primarily along the magnetic field lines, which would
in turn cause a more gradual decrease of $E_{p}$ with $\Psi$ (or $\alpha$).  This may be
a very plausible explanation - there may exist physical
situations which require either a higher magnetic field
or characteristic electron Lorentz factor, in which it is very efficient to accelerate
along the magnetic field lines.  On the other hand, the
electrons' Lorentz factors could be only mildly relativistic (instead
of $\sim 100$ as we assumed for the dashed line in Figure 4), which
would lead to a smaller relative decrease ($\sim 1/\gamma_{e}$) in $E_{p}$  as a function
of pitch angle. Finally, it is of course possible that
this model is incorrect and
an alternative explanation is needed to accommodate bursts above the
$\alpha=-2/3$ line of death.  Still, it is encouraging that the 
global distribuitons and observed trends are accomodated well by
our scenario.

  \begin{figure}[t]
  \includegraphics[height=.35\textheight]{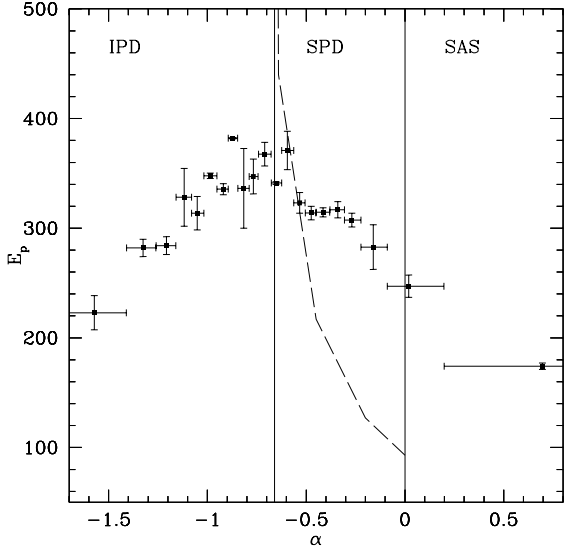}
  \caption{Observed $E_{p}$ vs. $\alpha$, for $\sim 2000$
  time resolved spectra.  Note the
  change in the sign of the correlation as $\alpha$ transistions
  from the IPD to SPD regime.}
\end{figure}

\subsection{Particle Acceleration and Remaining Questions}
   Our results bring to light the fact that particle acceleration in GRBs
is a quite poorly understood problem.  Usually, it is assumed
that the radiating particles in GRBs are accelerated via repeated
scatterings across the (internal) shocks.
  This mechanism, however,
predicts several features in the electron distribution not borne out
by the data.  First, it 
 has been shown \cite{kir00}
that these
repeated crossings of the shock result 
in a power law particle
distribution with a well defined index, $p = -2.23$, which would give a 
high energy synchrotron photon index $\beta$ of -1.62 (or -2.12 for the
``cooling'' spectrum, e.g. \cite{spn98}).
 Although this is consistent with some afterglows, this is
certainly is not true for many bursts in the 
prompt phase.
In our synchrotron models above, the high energy photon index 
$\beta=-(p+1)/2$, where $p$ is the high energy index of the emitting
particle distribution.  The parameter
$\beta$ can vary by a factor of
4 (or more!) throughout a single burst (see, e.g., \cite{pre99}),
 reflecting a huge variation (from $1$ to $9$) in the parameter
$p$ of the underlying
particle distribution - this is well beyond the statistical limits placed on
$p$ by shock acceleration simulations.\footnote{Preliminary results indicate 
that, instead, our adopted picture of stochastic acceleration can
produce the necessary electron distributions needed to explain the observed photon
spectra in at least the IPD synchrotron emission scenario. }
In addition, shock acceleration predicts an {\em isotropic}
distribution of electrons.  Our work 
 suggests that in a large fraction of GRBs, the
particle  acceleration is not isotropic but along the magnetic field lines.
Thus, there are many crucial open questions related to the physics of
particle acceleration in relativistic plasmas, and it is clear 
 that a complete
  investigation of this phenomenon in the context of GRBs
  is necessary.

\section{Prompt X-ray Emission}
 Although most GRBs
  emit the bulk of their prompt radiation 
  in the soft gamma-ray energy band (as their names imply),
 with the availability of recent X-ray observations of the
  GRB prompt emission from GINGA \cite{str98},
  Beppo-SAX \cite{hei01, fron00},   
  untriggered low energy BATSE data \cite{kip01}, and HETE-II \cite{bar01},
  an increasing number of GRBs with spectra that peak in the X-ray
  energy range ($\sim 1-40$ keV) have been discovered.  Current
  estimates \cite{kip01, hei01}, suggest that $\sim 30$\% of all bursts
  may have $E_{p}$ values below $\sim 40$ keV.   There has
  been considerable speculation over the possibility that  these
  low $E_{p}$ spectra may be a result of redshift effects - e.g. that
  these GRBs occur extremely high redshifts ($(1+z) > 5$), and therefore
  their observed spectra are shifted into the X-ray band.  However, as we show below, intrinsic
  properties of the GRB could also very well produce low peak energies.
  Below we discuss the role of both cosmological and intrinsic effects in
  producing X-ray rich GRBs.
  
   \subsection{Cosmological Effects}
   \begin{figure}[t]
  \includegraphics[height=.35\textheight]{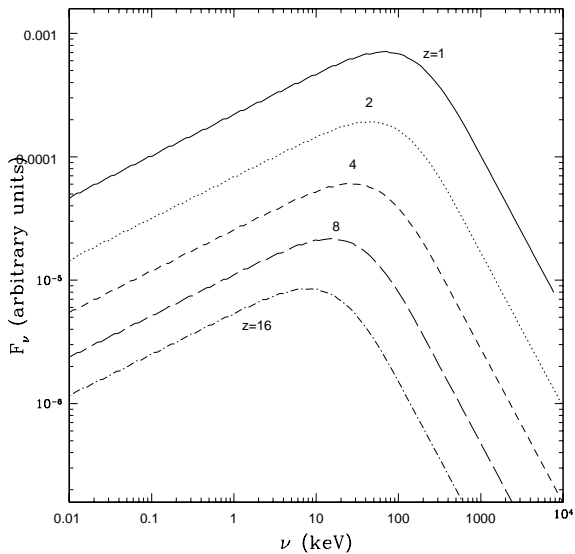}
  \caption{A sample $F_{\nu}$ spectrum as a function of redshift.}
  \end{figure}
  It is straightforward to ask: if we take a burst with
  a ``typical'' flux and peak energy at a redshift of $1$ and we redshift
  it until its peak energy falls in the X-ray band, {\em would the burst
  be detectable (by current instruments)?}. 
  Consider a burst at redshift $z$ with specific luminosity $L(\nu_{o})$ in the
  cosmological rest frame of the source.  The observed flux per unit
  frequency is 
  \begin{equation}
  f_{\nu} \propto \frac{L(\nu(1+z))}{d\Omega d_{metric}^{2} (1+z)}
  \end{equation}
  where $\nu=\nu_{o}/(1+z)$, $d\Omega$ is the geometric solid angle into which
  the GRB outflow is confined, $d_{metric}$ is the metric distance
  defined as 
  \begin{equation}
  d_{metric} = \int_{0}^{z}(c/H_{o})\frac{dz}{\sqrt{\Omega_{\Lambda} +
\Omega_{m}(1+z)^{3}}}.
\end{equation}
For our purposes, we assume a simple standard synchrotron spectrum (in
the IPD regime).
 A burst at a redshift of $1$ with $E_{p} = 200$ and
  a flux of $10^{-6} erg/cm^{2}/s$ would have an observed flux of about $5 \times 10^{-9}$
  when $E_{p}$ is redshifted to $30$ keV.  On the other hand, for a burst with
  $E_{p} = 100$ keV and a flux of $10^{-5} erg/cm^{2}/s$ at $z=1$,
  it is easily detectable (flux $= 10^{-8}$ erg/cm$^{2}$/s) out
  to a redshift of $10$, where $E_{p} = 10$ keV.  
  We note that the Heise BeppoSAX sample \cite{hei01} has fluxes between $10^{-7}$
  and $10^{-8}$ erg/cm$^{2}$/s, while the fluxes from Kippen's
  BATSE sample \cite{kip01, kip00} ranged from $10^{-7}$ to $5 \times 10^{-9}$ erg/cm$^{2}$/s.
   Unfortunately, because of the broad GRB luminosity function 
   and intrinsic $E_{p}$ distribution (i.e. because there are no
  standard intrinsic luminosity and $E_{p}$ values for all GRBs) all
  we can say at this point is that it is {\em possible} that these
  low $E_{p}$ bursts are very high redshift bursts.  We do note that the highest fluxes in
  both data samples require the bursts to be on either the high end
  of the intrinsic luminosity distribution or the low end of
  the intrinsic $E_{p}$ distribution in order to be consistent with the redshift
  interpretation.  Nonetheless, we can neither rule out nor strongly favor
  the high redshift interpretation of these bursts.
  \begin{figure}[t]
  \includegraphics[height=.35\textheight]{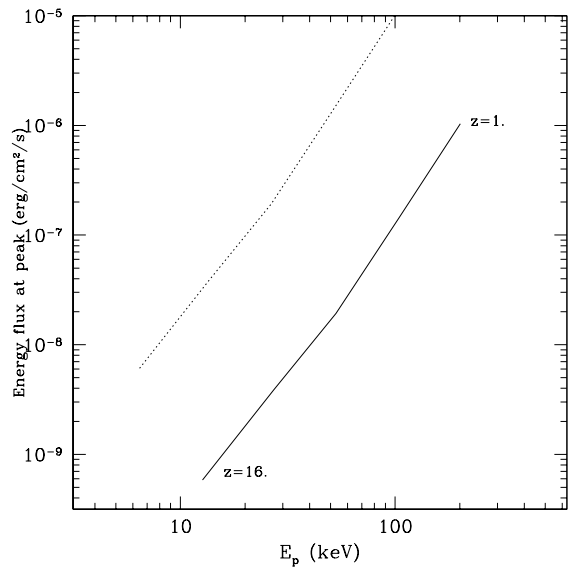}
  \caption{Energy flux at the peak of the $F_{\nu}$ spectrum vs.
  $E_{p}$ for changing redshifts of the burst, given two initial
  ``starting'' values of $E_{p}$ and flux at $z=1$.}
  \end{figure}
  
  \subsubsection{Luminosity Evolution}
  There have been some suggestions that GRB luminosity function evolves
  with redshift, in the sense that bursts at high redshift
  tend to have intrinsically higher luminosity \cite{lr01}. 
  {\em This statement
  is based on those bursts with redshifts and luminosities from the
  luminosity-variability relation} \cite{frr00, rei00}, and so caution should be exercised
  until this relation is more definititively shown to provide valid
  redshifts.  However, if we do adopt the redshifts and luminosities from
  the L-V relation, then there is significant evidence that GRBs
  are brighter at higher redshifts. This effect allows
  for an even greater probability that very high redshift bursts are detectable
  by current instruments.
  For example, \cite{lr01} find the average GRB luminosity
  $L \sim (1+z)^{1.4 \pm 0.5}$.  Therefore, the observed flux  decreases
  less rapidly as a function of redshift, than if there were no luminosity
  evolution.  
  \footnote{However, we mention one important caveat:
   If whatever causes the GRB luminosity function to
  evolve also causes $E_{p}$ to evolve, then there would be a decreased
  dependence of $E_{p}$ on redshift and one would have to place the burst
  at even higher redshifts
  to shift $E_{p}$ into the X-ray band.}  
  
   \subsection{Intrinsic Effects}
   The analysis above considered how a particular burst behaves as
   it is moved out in redshift space, for its particular set of observed
   properties.  However, as has been suggested repeatedly and confirmed
   by bursts with measured redshifts,  
  GRB intrinsic properties vary substantially from burst to burst.  In some
  cases, the intrinsic properties can act in such a way as to mimic redshift effects. 
  
  For example, again assuming a standard synchrotron spectrum (which
  we do to be concrete, but which is not necessary), we can investigate
  how the $\nu F_{\nu}$ flux changes as a function of various physical
  parameters.  We take the magnetic
  field as an example.   Figure 7 shows how the spectrum shifts
  with decreasing magnetic field from the $\gamma$-ray to
  X-ray regime.  
  The flux at the peak is approximately proportional to 
  $B^{2}$ while the peak energy goes as $B$.  Figure 8 shows
  the peak flux as a function of $E_{p}$.    
  Here we
  see that changing the magnetic field could also very reasonably
  produce X-ray bursts.  Other possibilities include decreasing
  the bulk Lorentz factor, or changing the electron spectrum
  (such as the density or minimum energy of the electrons).

    \begin{figure}[t]
  \includegraphics[height=.35\textheight]{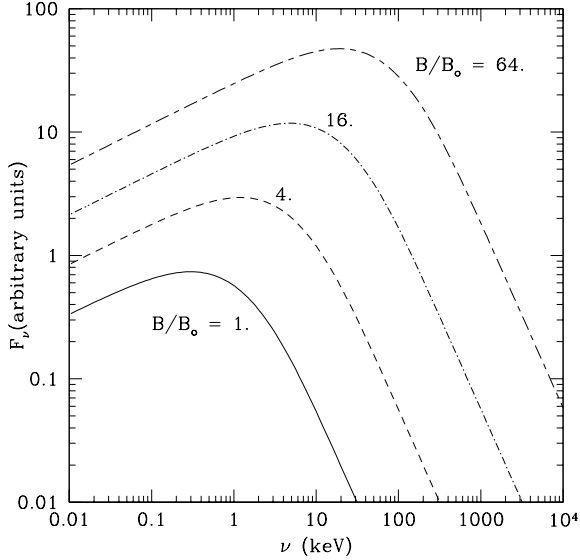}
  \caption{A synchrotron $F_{\nu}$ spectrum as a function
  of magnetic field.}
  \end{figure}
  There are in fact several pieces of evidence suggesting that
  intrinsic effects are playing the dominant role in producing
  the low $E_{p}$ bursts:

  \begin{figure}[t]
  \includegraphics[height=.35\textheight]{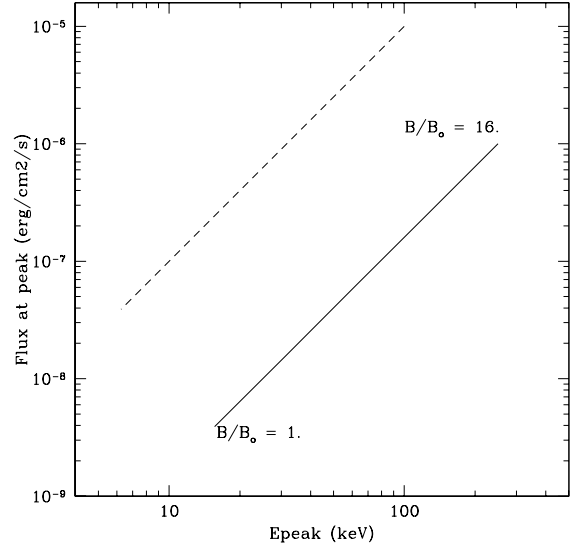}
  \caption{Same as Figure 6, but for a changing magnetic field.}
  \end{figure}
  
  {1. One clue comes from the fact that the Kippen et al. bursts \cite{kip01, kip00}, 
   fall along an extension of the
  well known observed hardness-intensity correlation \cite{mal96} 
  in BATSE GRBs (see Figure 3 of \cite{kip00}).  
  Lloyd, Petrosian, \& Mallozzi \cite{lpm00} found that this correlation cannot
  be produced by redshift effects alone, given any reasonable (i.e. not a $\delta$-function) 
  GRB luminosity function.
  That is, even for a fairly narrow intrinsic peak energy distribution,
  the observed hardness-intensity correlation cannot be due to cosmological expansion
  (e.g. lower flux and lower $E_{p}$ do not necessarily 
  imply higher redshifts because the broadness of the GRB luminosity
  function washes out this effect).  In fact, they found that a correlation
  between the intrinsic peak energy and the emitted energy of the GRB would
  naturally reproduce the observed correlation, and that such a relation between
  peak energy and total energy/luminosity is natural in (but not limited to) a synchrotron 
  emission model.}

  {2.
  The observed durations
  of the XRBs are not unusually long or smooth.  
  At the highest redshifts (e.g. $1+z\sim 10$), the duration and pulse width are
  dilated by a factor of $10$.  Although there exists no standard in GRB temporal
  properties, one might expect that if these bursts were at very high
  redshifts, their time profiles would be particularly long or smooth - this
  does not seem to be the case.
  Heise et al. \cite{hei01} also show that
  the durations of the bursts in his sample are completely consistent
  with the global population and not unusually long. 
  }
  
 {3. The Kippen and Heise sample of low $E_{p}$ bursts are based on the {\em average}
 (over the duration of the burst) spectral properties of the burst.
 However, there have been some time resolved spectral analyses in
 which  observations of 
  X-ray rich {\em pulses} within a burst are reported (e.g. \cite{smi01, fron00}).
  For at least these bursts, the X-ray ``richness'' of the pulses {\em must}
  therefore be coming entirely from intrinsic effects.  }

\section{Prompt Optical and Very High Energy Emission}

\underline{\bf Optical} There are several mechanisms in a GRB that may
produce prompt emission at optical wavelengths. And in fact in
one burst - GRB 990123 - such emission was seen as
a bright optical flash ($m_{v} \sim 9$)
during the prompt phase \cite{ak99}. The most common
interpretation of this optical flash is that it resulted from
the blast wave reverse
shock propagating
into the dense GRB ejecta. Because the temperature 
in this region is smaller relative to the shocked ambient medium
by  a factor of $\sim \Gamma^{2}$,
the emission from this interaction peaks in the optical energy band.  
Certain conditions (sensitive to $\Gamma$)
are required to actually produce such a flash (see \cite{sp99, sod01} for discussion);
if this interpretation is correct, it offers the possibility of constraining - among
other things - the bulk Lorentz factor and baryonic content of  the GRB outflow.
 
  An alternative interpretation of this bright optical
  flash involves the interaction of the GRB
 radiation front \cite{bel01, tm00, mrrr01} with the ambient medium.
In this scenario,  the $\gamma$-ray photons travelling ahead of the blast wave are side-scattered
by the ambient medium.  These photons 
then pair produce with the forward streaming photons, and pair enrich the medium.
The available energy must distributed among these pairs, and therefore
the peak of the initial afterglow emission is lower in energy (relative to
the standard picture without pairs)
by a factor of  $(m_{e}/m_{p})^{2}$. As a result, the very early GRB afterglow (possibly
coincident with the prompt emission) is brightest
in the optical.  In order for this mechanism to work
efficiently, the medium is constrained to a particular density
profile (see \cite{bel01} for more details).  Hence,
if this mechanism succeeds, the prompt optical GRB emission affords
us the opportunity to constrain the surrounding GRB environment.  
 
\underline{\bf Very High Energy}
  If synchrotron emission is indeed responsible for the prompt
  GRB emission in the low energy gamma-ray band (see above), then
  we expect an inverse-Compton component from this emission.
  The inverse Compton component is boosted by a factor of $\Gamma^{2}$
  and therefore peaks in the $\sim 10$ GeV range.  There is also the
  possibility of observing very high (i.e. TeV) emission from a high
  energy proton component \cite{tot99}, as well as other high energy
  phenomena  (such as neutrinos; see \cite{wax00} for a discussion). A number of upcoming instruments
  (GLAST, Agile, Milagro, etc) will be sensitive in the range
  from 10's of MeV to TeV energies and have the potential of shedding
  much light on this relatively unexplored  but important energy range of
  of GRB emission.
  
\section{Conclusions}
   We have discussed the prompt emission of GRBs, focusing
   particularly on the gamma-ray and X-ray emission properties.  We find 
   that a generalized synchrotron
   emission model does a good job of explaining the behavior of the
   prompt spectra in the $\sim 20$ keV  $- 1$ MeV range.  Moreover, we find that 
   although the increasing number of low $E_{p}$ ($< 40 keV$) GRBs may
   be consistent with a high redshift interpretation, the data
   appear to suggest that in fact variations in the bursts' intrinsic
   properties (e.g. magnetic field, electron energy density, etc.) are probably
  responsible for producing these ``X-ray rich'' bursts.
    We emphasize the importance
   of broadband (from eV to TeV) observations of prompt emission in order to
    gain a complete 
   understanding of the physics of Gamma-Ray Bursts.

\begin{theacknowledgments}
I would like to thank the organizers for an
interesting and stimulating conference.  I would also like
to thank Vahe' Petrosian, with whom much of the work on
the prompt gamma-ray emission was done.
\end{theacknowledgments}


\doingARLO[\bibliographystyle{aipproc}]
          {\ifthenelse{\equal{\AIPcitestyleselect}{num}}
             {\bibliographystyle{arlonum}}
             {\bibliographystyle{arlobib}}
          }

\end{document}